\documentclass{jfm}

\usepackage{epstopdf}
\usepackage{graphicx}
\usepackage{dcolumn}
\usepackage{bm}
\usepackage{url}
\usepackage{hyperref}
\usepackage{color}
\usepackage{comment}
\usepackage{natbib}
\usepackage{gensymb}
\usepackage[utf8]{inputenc}
\usepackage[T1]{fontenc}
\usepackage{siunitx}
\usepackage{longtable}
\usepackage{amsmath}
\usepackage{amssymb}
\usepackage{mathrsfs}
\usepackage{commath}
\usepackage{esint}
\usepackage{epsfig}
\usepackage{rotating}
\usepackage{float}
\usepackage{array}
\usepackage{booktabs}
\usepackage{caption}

\title{How shape and flapping rate affect the distribution of fluid forces on flexible hydrofoils}

\author{Paule Dagenais and Christof M. Aegerter\corresp{\email{aegerter@physik.uzh.ch}}}
\affiliation{Physik-Institut, University of Zurich, Winterthurerstrasse 190, 8057 Zurich, Switzerland}

\begin{document}

\maketitle

\begin{abstract}

We address the fluid-structure interaction of flexible fin models oscillated in a water flow. Here, we investigate in particular the dependence of hydrodynamic force distributions on fin geometry and flapping frequency. For this purpose, we employ state-of-the-art techniques in pressure evaluation to describe fluid force maps with high temporal and spatial resolution on the deforming surfaces of the hydrofoils. Particle tracking velocimetry (PTV) is used to measure the 3D fluid velocity field, and the hydrodynamic stress tensor is subsequently calculated based on the Navier-Stokes equation. The shape and kinematics of the fin-like foils are linked to their ability to generate propulsive thrust efficiently, as well as the accumulation of external contact forces and the resulting internal tension throughout a flapping cycle.

\end{abstract}

\maketitle

\section{Introduction}

The interplay between fins shapes, elastic properties, hydrodynamic forces, passive and controlled kinematics is the subject of persistent and active research \citep{Videler1975, Geerlink1986, Tangorra2007, Lauder2015, Puri2018a}. Replicating the flexible fins or body of fish using hydrofoils with controlled motion programs has proven a powerful tool to investigate the kinematics and propulsive forces of swimmers \citep{Shelton2014}. Methods to quantify the fluid velocity field are typically based on imaging tracer particles seeded into the flow (PIV/PTV) \citep{Maas1993, Dracos1996, Raffel1998, Pereira2006}. The vortex wakes and thrust production of flapping foils with various geometries and flexibility have been extensively characterized in the literature \citep{Triantafyllou2004, Godoy-Diana2008, Bohl2009, Kim2010, Green2011, David2012, Marais2012, Shinde2014, David2017, Lucas2017, Muir2017}. Furthermore, many researchers have resorted to particle velocimetry experiments in order to quantify the flow field of aquatic animal appendages and bioinspired synthetic fins \citep{Blickhan1992, Stamhuis1995, Muller1997, Drucker1999, Lauder2000, Muller2000, Drucker2001, Muller2001, Nauen2002a, Nauen2002b, Drucker2005, Muller2006, Tytell2006, Lauder2007a, Tangorra2007, Muller2008, Tytell2008, Tangorra2010, Flammang2011a, Flammang2011b, Dewey2012, Esposito2012, Ren2016a, Ren2016b, Mwaffo2017}.

The research field making use of PIV/PTV-based pressure evaluation covers a large collection of hydrodynamics problems, from micro channel junction flow and turbine blades to animal locomotion \citep{Gresho1987, Jakobsen1999, Baur1999, Gurka1999, Fujisawa2006, Liu2006, vanOudheusden2006, Murai2007, vanOudheusden2007, vanOudheusden2008, Windsor2008, Jardin2009, Lorenzoni2009, Khodarahmi2010, deKat2012, Ragni2012, deKat2013, Panciroli2013, vanOudheusden2013, Dabiri2014, Joshi2014, Tronchin2015, Lucas2017, McClure2017, Mwaffo2017}. Accurate and non-invasive methods to measure the fluid forces directly on the surface of flapping fins are essential to investigate the mechanisms of underwater propulsion. In the present work, we combine fluid velocity measurements from 3D particle tracking velocimetry with hydrodynamic stress calculations based on the Navier-Stokes equation to obtain well resolved dynamic maps of fluid forces on the surface of flexible fin-like foils operating in different flow regimes. We address the long-standing question of how the geometry and kinematics of oscillating hydrofoils influence their spatio-temporal distributions of hydrodynamic forces. We focus on the effects of flapping frequency, width, length and edge curvature (straight versus bilobed). This approach offers a framework to investigate how specific morphological and kinematic features can constitute an advantage for certain hydrodynamic functions. The implications of this study reach into the fields of animal aquatic locomotion, as well as the engineering design of biomimetic underwater vehicles. Such propulsive systems are gaining importance due to their potential applications in the monitoring, maintenance and exploration of underwater environments \citep{Low2006, Lauder2007b, Low2007, Zhou2012, QingPing2013}.

\section{Material and Methods}

\subsection{Particle tracking velocimetry}

We use a three-dimensional, three components particle tracking velocimetry approach (3D-3C PTV) to quantify the flow generated by the synthetic fins. Technical aspects of three-dimensional PTV are well described in \citet{Pereira2006}. The basic principle of PTV relies on tracer particles seeded in the flow, illuminated by a laser beam and imaged at regular time intervals to track the position of each particle, allowing the subsequent reconstruction of the fluid velocity field. Because the aim is to calculate the hydrodynamic stress tensor from the velocity fields, this experimental method presents several benefits compared to other techniques such as tomographic or scanning stereo-PIV, the most outstanding one being the possibility to instantaneously capture the whole flow volume, allowing for the straight-forward reconstruction of the 3D velocity vectors everywhere inside that volume.

\begin{figure}
\centering
\includegraphics[width=10cm]{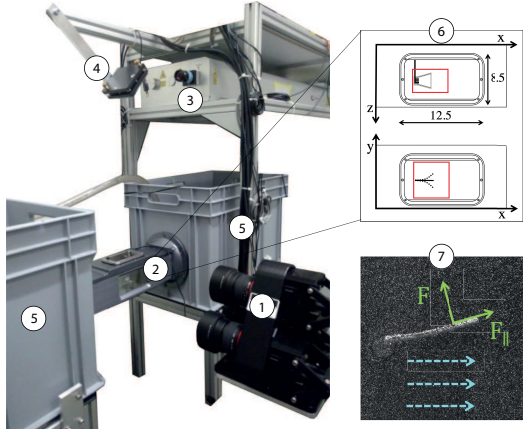}
\caption{Experimental set up: (1) camera triplet (4 MP, 85 mm lenses, sensor size 11.3$\times$11.3 mm$^2$, magnification 0.3, max. frequency 180 Hz) arranged in a triangular configuration on a plate located at $\sim$465 mm from the middle of the (2) flow chamber, illuminated by a (3) laser beam (double-pulsed, Nd:YAG, wave-length 532 nm, max. 120 mJ/pulse) expanded by a pair of cylindrical lenses and deflected by (4) a mirror. (5) recirculating system composed of two tanks (total volume of $\sim$160 L) and a pump connected to a pipe. (6) sketch of the flow chamber with inner dimensions (in mm), top view ($x-z$ plane) and frontal view ($x-y$ plane), measurement volume ($\sim$50$\times$50$\times$20 mm\textsuperscript{3}) indicated with red rectangles, enclosing the trapezoidal fin. (7) example of an image captured by one of the three cameras, with tracer particles and hydrofoil midline clearly visible (direction of the free-stream indicated by blue dashed arrows and direction of the normal and tangential forces indicated by green arrows).}
\label{f:set_up}
\end{figure}

The different components of the experimental set up are shown in figure \ref{f:set_up} with their technical specifications. The working fluid is water and the seeding tracers are polyimide particles with a diameter of $\sim$50 $\mu$m, which have been used in previous particle velocimetry experiments to investigate the flow fields generated by fish fins \citep{Flammang2011a}. The measurement volume in our experiments is 50$\times$50$\times$20 mm\textsuperscript{3}, with an approximate number of 8$\times$10\textsuperscript{4} seeding particles inside that interrogation domain. Three cameras mounted on a plate in a triangular arrangement are used to record the 3D fluid velocity vectors $\vec{u}=(u_{x},u_{y},u_{z})$. The flow chamber has transparent walls on three sides and a fixation wall on one side for inserting the fin model, actuated with a servomotor fixed outside the chamber. The hydrofoils are mounted horizontally inside the tunnel, with their rotation axis parallel to the $z$ axis and perpendicular to the cameras plate. A frontal perspective would make the detection of the particles more arduous in front of the fin surface. Water tanks are connected to both sides of the flow chamber in a recirculating system, and a pump is pushing water inside a pipe from one tank to the other to control the upstream fluid velocity. A flow straightener is installed at the inlet of the tunnel to insure laminar incoming stream. All the experiments were performed using the V3V-9800 system (TSI Incorporated, 500 Cardigan Road, Shoreview, Minnesota 55126 USA), which is characterized in \citet{Lai2008}. The laser double pulse timing and the cameras capturing frequency are synchronized in a method called frame straddling: the time difference between a pair of position fields is determined by the time difference between a pair of laser pulses ($\delta t = 2.5$ ms), and the rate at which velocity fields are recorded (80 Hz) is half the cameras acquisition frequency, yielding a time separation of $\Delta t$=12.5 ms between the velocity fields. The reconstruction of the particles positions is based on 2D Gaussian fit of the particles intensity distributions and a triplet search algorithm is used for the 3D positions fields. The velocity vectors are computed by tracking the particle displacements between subsequent laser pulses, using a relaxation method to achieve a probability-based matching. These processing steps are conducted using the V3V software (version 2.0.2.7). An appropriate combination of median filter, velocity range and smoothing filter was applied to the raw velocity fields to reduce the noise level and remove the worst outliers. Moreover, a mask is applied over the hydrofoil during image processing to avoid detecting ghost particles inside its boundary. Finally, the raw vectors are interpolated on a regular 3D grid using Gaussian interpolation, yielding a final spatial resolution of 0.75 mm in each direction. Temporal and spatial resolutions were chosen along recommendations offered in previous studies where the flow velocity field was used to compute hydrodynamic pressure \citep{deKat2012, Dabiri2014, Wang2017}. The grid points inside the hydrofoil boundary, where no particle is detected owing to the mask, are not attributed a velocity value and are not involved in the hydrodynamic stress calculation. The closest grid points to the real fin boundary define a virtual boundary, where the surface distributions of hydrodynamic forces are evaluated.

The foil midline is visually tracked over time in the $x$-$y$ plane (see panel 7 of figure \ref{f:set_up}). For each time frame, points are manually superimposed on the fin midline and fitted with a polynomial of degree 2. The 3D hydrofoil is reconstructed based on that fitted quadratic curve, assuming that deflection occurs only along one axis. The virtual object has a larger width than the actual foil (0.725 mm away from the real surface on each side) owing to the fact that the particles can not be resolved directly at the fluid-solid interface. In order to evaluate if the reconstructed foil boundary is sufficiently close to the real surface where we want to extract the hydrodynamic pressure, we used the criterion that the virtual surface should be located within the fluid boundary layer. Based on the assumption of a laminar boundary layer, its thickness can be expressed as (in the simplified case of a flat plate) \citep{Prandtl1952}:

\begin{equation}
\delta=5 \sqrt{\frac{\nu x}{U}}
\label{eq:BL}
\end{equation}

We could estimate that the thickness of the fluid boundary layer is >1 mm everywhere on the fin except at the most proximal region (<10\% of the fin length). The virtual surface thus remains inside the boundary layer of the real object. Even though the boundary layer of an oscillating fin involves more complex phenomena such as turbulence, separation and reattachment \citep{Obremski1967, Kobashi1980, Arnal1984, Incropera2007, Kunze2011}, these flow effects would tend to make the boundary layer extend further away from the solid surface. Therefore, the assumption of laminarity of the boundary layer gives a conservative estimate of the boundary layer width and thus allows us to verify that the virtual foil boundary provides an accurate representation of the real surface distributions. This point was verified in a previous study \citep{Dagenais2019}, where we used a control volume analysis to compute the forces generated by a flapping fin in similar conditions, and showed that the results compared favorably to the integrated force distributions over the fin surfaces.

\subsection{Morphologies and kinematics of the fin models}

\begin{figure}
\centering
\includegraphics[width=9cm]{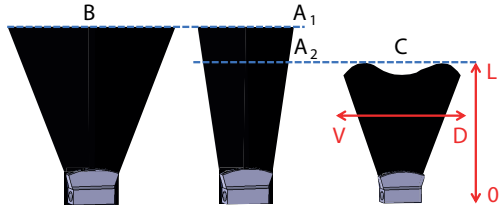}
\caption{Four fin geometries under study. Principal axes of the fin are shown in red: proximo-distal axis (from 0 to $L$) and dorso-ventral axis (from V to D).}
\label{f:fin_models}
\end{figure}

\begin{figure}
\centering
\includegraphics[width=12cm]{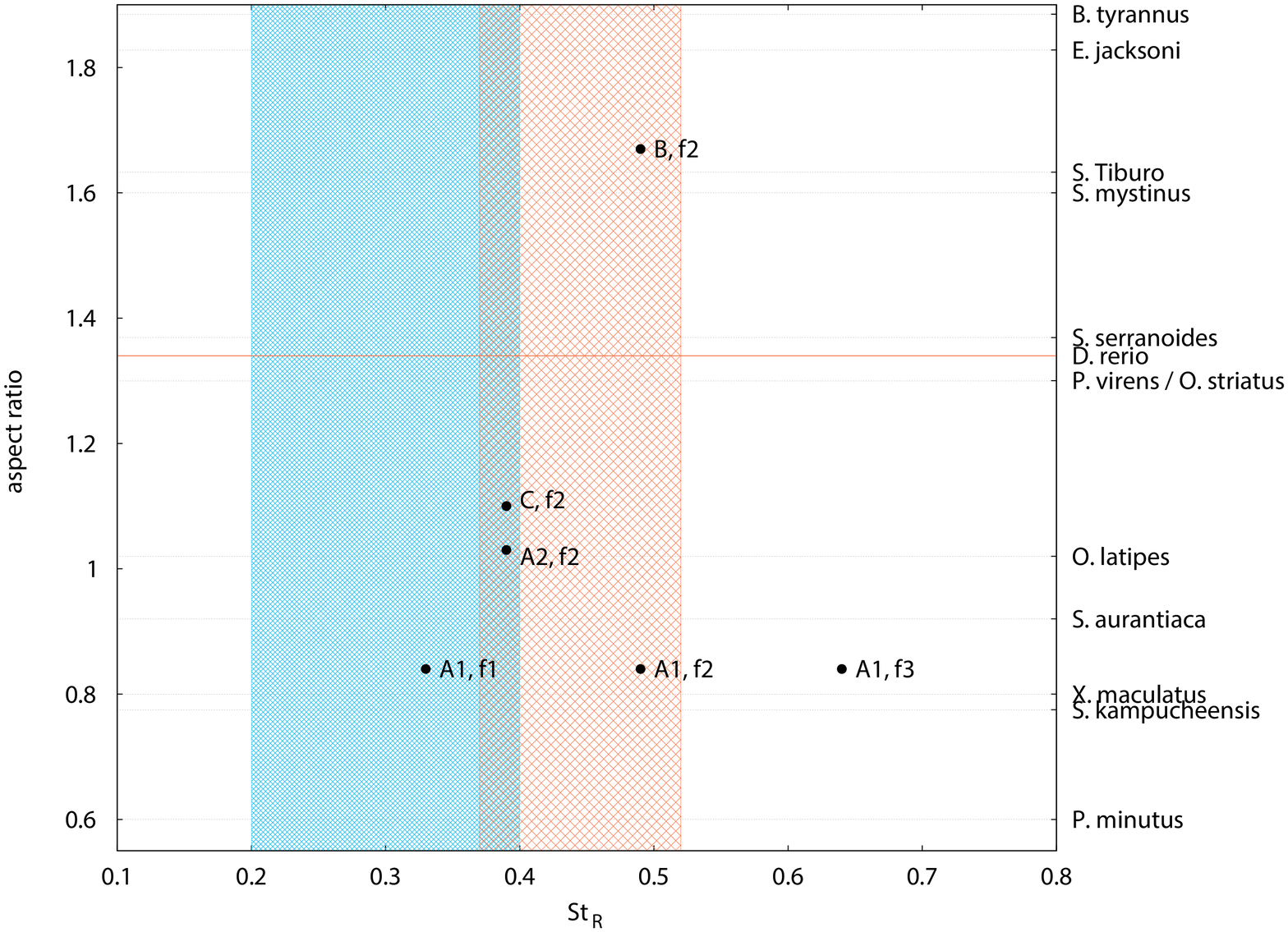}
\caption{Black dots: aspect ratio and \textit{rigid} Strouhal number ($St_r$, calculated based on the rigid projection of the peduncle) for each fin model. Horizontal lines: caudal fins aspect ratios for various fish species, derived from \citet{Sambilay2005, Offen2008, Parichy2009, Naruse2011, Plongsesthee2012, Bohlen2016, Kottelat2017}. Dashed blue zone: range of Strouhal numbers typically associated with efficient propulsion for flapping foils \citep{Triantafyllou2000, Taylor2003}. Dashed pink zone: range of Strouhal numbers for the adult zebrafish caudal fin, derived from \citet{Parichy2009, Palstra2010, Mwaffo2017}}
\label{f:rigid_St_AR}
\end{figure}

The synthetic fins are illustrated in figure \ref{f:fin_models} and characterized in table \ref{t:exp_parameters} and figure \ref{f:rigid_St_AR}. Taking shape A\textsubscript{1} as the reference geometry, A\textsubscript{2} constitutes a shorter version and shape B, a wider version. Shape C has the same length as shape A\textsubscript{2}, but presents a bilobed trailing edge. The aspect ratio (AR) of a fin is defined as the square of the span (width at largest point) divided by the area. Even though the fin models do not mimic any fish species in particular, their geometries can be compared to caudal fins with low aspect ratios (AR=0.84 for shape A\textsubscript{1}, 1.03 for shape A\textsubscript{2}, 1.67 for shape B and 1.1 for shape C). Examples include the \textit{Schistura} genus, the \textit{Oryzias} genus, the zebrafish (\textit{Danio rerio}) and the platy fish (\textit{Xiphophorus maculatus}), as shown in figure \ref{f:rigid_St_AR} \citep{Sambilay2005, Offen2008, Parichy2009, Naruse2011, Plongsesthee2012, Bohlen2016, Kottelat2017}.

To produce the foils, a rigid cast of their negative form was 3D printed, then liquid PDMS (polydimethylsiloxane) was poured inside the cast and the supporting rod was inserted at the base. The material was cured for 36 hours at 58$^{\circ}$C. The resulting flexible membranes have a thickness of 0.55 mm. The properties of the cured PDMS can be found in the MIT material property database (\url{http://www.mit. edu/~6.777/matprops/pdms.htm}). Most importantly, the mass density matches that of water. A cantilever deflection set up was employed by Sahil Puri (University of Zurich) to characterize the elastic properties (see \citet{Puri2018a, Puri2018b} for details), yielding a value of 0.8 MPa for the Young's modulus. An external velocity of $u_{\infty}=55$ mm/s is imposed in all experiments, and the fins are actuated at their leading edge with a sinusoidal pitching motion with angular amplitude $\theta_{0}=11^{\circ}$ in all cases. Shape A\textsubscript{1} was selected to test the effects of pitching frequency, using $f\textsubscript{1}$=1.9 Hz, $f\textsubscript{2}$=2.8 Hz and $f\textsubscript{3}$=3.7 Hz. The natural frequency of this fin model (first mode) was estimated by releasing it from a rest position and measuring the decaying oscillation of its tip (in water). A value of 2.4 Hz was found, in between the lowest and intermediate frequencies tested. For the experiments involving shapes A\textsubscript{2}, B and C, frequency $f\textsubscript{2}$=2.8 Hz is used.

Two dimensionless parameters allow to characterize the flow regime of the hydrofoils. The Strouhal number ($St$) encompasses the propulsion dependence on tail oscillation. The Reynolds number ($Re$) describes the viscous versus inertial effects and determines the transition from a laminar to a turbulent flow. These parameters are based on the kinematic viscosity of water ($\nu$), the external fluid velocity ($U=u_{\infty}$), the foil length ($L$), the flapping frequency ($f$) and the tip excursion amplitude ($A$):

\begin{equation}
Re=\frac{U \cdot L}{\nu} \hspace{1cm}  St=\frac{f \cdot A}{U}
\label{eq:ReSt}
\end{equation}

Caudal fins found in nature are extremely diverse and cover a vast range of flow parameters, depending on species but also on developmental stage and behavior. The swimming velocities catalogued in the literature usually correspond to the maximal speeds which fish can sustain for a short time only, yielding $Re$ from 10\textsuperscript{4} to 5$\times$10\textsuperscript{6} (using the animal body length) \citep{Wardle1975, Yates1983, Vogel1994}. These high values are not representative of slower gaits at which fish can also operate in natural conditions \citep{Bainbridge1960, Weerden2013}. For example, based on the vast literature about zebrafish hydrodynamics, it can be estimated that this species operates at $Re$ in the range [390-2600] (using the caudal fin length to allow direct comparison with isolated hydrofoils) \citep{Parichy2009, Palstra2010, Mwaffo2017}. In the present study, $Re$ is situated within [1062-1375] (table \ref{t:exp_parameters}). Although not characteristic of fast bursts or maximal velocity locomotion, this flow regime can be compared to the case of fish cruising at roughly 0.5 to 0.9 body lengths per second, pertinent for certain species \citep{Sambilay2005} and specific behaviors such as foraging, chemotaxis and exploration.

Furthermore, studies have demonstrated that fundamental flow features of flapping fin propulsion can be captured by experiments and simulations performed at $Re$ lower than values typically measured for fast swimming fish, down to the order of 10\textsuperscript{3} \citep{Lauder2005, Bozkurttas2006, Buchholz2006, Kern2006, Mittal2006, Bozkurttas2009, Liu2017}. Indeed, for sufficiently high $Re$ ($\geq$ 10\textsuperscript{3}), this parameter plays a minor role compared to $St$ in defining the flow topology \citep{Lentink2008, Green2011}.

In the case of flexible foils, $St$ depends on the tip excursion amplitude and has to be measured rather than fixed prior to the experiments. Therefore, we additionally define the \textit{rigid} Strouhal number ($St_r$), an input parameter based on the excursion amplitude of an equivalent rigid fin (same length and angular amplitude at the base). For the six experimental cases, we obtain $St_r$ in the range [0.33-0.64] (figure \ref{f:rigid_St_AR}). Strouhal numbers between 0.2 and 0.4 usually yield the highest propulsion efficiency \citep{Triantafyllou2000, Taylor2003}. Nevertheless, wider ranges of $St$ within [0.2-0.7] were reported in various fish species and for different developmental stages \citep{Eloy2012, Eloy2013, Weerden2013, Xiong2014, vanLeeuwen2015, Link2017}. For example, adult zebrafish flap their caudal fins with $St$ in a window of [0.37-0.52], depending on their swimming mode \citep{Parichy2009, Palstra2010, Mwaffo2017}.

Aside from their pertinence in the fish world, the experimental parameters (AR, $Re$ and $St$) are similar to those selected in previous research about fluid-structure interactions of flapping foils \citep{Lauder2005, Godoy-Diana2008, Dai2012, Marais2012, Dewey2013, Quinn2014, Shinde2014, Quinn2015, David2017, Floryan2017, Liu2017, Rosic2017, Zhu2017}. We explore a specific portion of the flow parameters space to understand how propulsive efficiency and surface distributions of hydrodynamic forces depend on shape and frequency at relatively low Reynolds numbers, a topic of high interest in the field of fin propulsion and for the practical design of underwater vehicles relying on bio-inspired undulating membranes.

\begin{table}
\begin{center}
\begin{tabular}{c c c c c c c c c c}
shape                        & $L$  & $d$    & $Re$ \\
                                 & (mm) & (mm) &           \\
\hline
A\textsubscript{1}     & 25     & 14      & 1375  \\
A\textsubscript{2}     & 20     & 13      & 1100   \\
B                               & 25     & 24      & 1375  \\
C                              & 20      & 13      & 1062
\end{tabular}
\caption{Parameters of the fin models: length ($L$), width at the tip ($d$, distance between lobes tips for shape C), Reynolds number ($Re$) and \textit{rigid} Strouhal number ($St_r$).}
\label{t:exp_parameters}
\end{center}
\end{table}

\subsection{Calculations}
\label{sst:calculations}

Based on the PTV-velocity fields, the total hydrodynamic stress tensor \bm{$s$} is calculated at each node in the 3D domain, which is the sum of the scalar pressure field $p$ (multiplied with the identity matrix and a factor -1) and the viscous stress tensor \bm{$\tau$}:

\begin{equation}
s_{ij}=-p\delta_{ij}+\tau_{ij}
\label{eq:total_stress}
\end{equation}

The viscous stress tensor depends only on the spatial derivatives of the velocity field (where $\mu$ is the dynamic viscosity):

\begin{equation}
\tau_{ij}=\mu\bigg( \frac{\partial u_i}{\partial x_j}+\frac{\partial u_j}{\partial x_i} \bigg)
\label{eq:viscous_stress}
\end{equation}

Although it was included in the calculation, the viscous stress is much smaller than the pressure in the present case. Indeed, the fins operate in the inertial flow regime ($Re$>1000), although theoretically very close to the transitional range 300<$Re$<1000 \citep{McHenry2005}, where the normal stress component (dominated by the pressure) is typically larger than the viscous tangential stresses by at least two orders of magnitude.

The pressure evaluation is based on the Navier-Stokes equation \citep{Aris1990, Whitaker1968}:

\begin{equation}
\rho \frac{D\vec{u}}{Dt}=-\vec{\nabla}p+\vec{\nabla} \cdot \bm{\tau}+\rho \vec{g}
\label{eq:Navier-Stokes}
\end{equation}

The last term on the right corresponds to any type of body force such as gravity; it is included in the pressure term and omitted in the remaining development. The left side of the equation contains the material acceleration $D/Dt$. This term is evaluated in the Lagrangian frame, namely in the reference frame of the advected particle \citep{Dabiri2014}. Each component of the pressure gradient can be calculated from the Navier-Stokes equation:

\begin{equation}
\frac{\partial p}{\partial x_i}= \bigg(-\rho \bigg( \frac{\partial}{\partial t}+\sum_j{u_j \frac{\partial}{\partial x_j}}\bigg)
+\mu \sum_j{\frac{\partial^2}{\partial x_j^2}} \bigg) u_i \\
\label{eq:pressure_gradient}
\end{equation}

Instantaneous pressure fields are reconstructed through direct integration of the above equation with appropriate boundary conditions. We performed the pressure calculation using the \textit{queen2} algorithm from \citet{Dabiri2014}, available at \url{http://dabirilab.com/software/}. A null pressure value is assumed on the external boundary of the domain, in the undisturbed flow. The validity of that assumption relies on the conditions $H/D \geqslant 2$, where $H$ is half of the domain size \citep{Dabiri2014}. Substituting the tip excursion amplitudes for the characteristic dimension $D$, we conclude that our experiments lie just above that limit. For each node inside the domain, the pressure gradient is integrated along eight different paths (horizontal, vertical or diagonal) originating on the outside contour. The median from the eight resulting pressure values is finally selected. This algorithm offers the advantage of reasonable computation time even for large domains. The crucial step in the pressure calculation lies in the determination of the material Lagrangian acceleration. In the \textit{queen2} algorithm, a so-called pseudo-tracking scheme is applied. The position of a particle at an instant $t_{i+1}$ is approximated based on its initial location $\vec{x}_p(t_{i})$ and the velocity evaluated at this location, averaged between instants $t_{i}$ and $t_{i+1}$. The velocity of the particle at its estimated forward position at time $t_{i+1}$ is then employed to evaluate its acceleration at time $t_{i}$. The pressure gradient is evaluated in a quasi-3D manner, using in-plane velocity derivatives only, but combining integration paths in both the $x$-$y$ and the $x$-$z$ planes. The details of the calculation are presented in \citet{Dabiri2014}.

The total hydrodynamic stress tensor (\ref{eq:total_stress}) is projected on the surface of the solid object to obtain a stress vector, which is the total force per unit area generated by the fluid on the foil. Each $i$-component of the hydrodynamic stress vector acting on a surface with a unit normal vector $\hat{e}_{n}=(n_x,n_y,n_z)$ (oriented outwards) is expressed as:

\begin{equation}
S_i(\vec{x},t,\hat{n})=s_{ij}(\vec{x},t) \cdot n_j
\end{equation}

In section \ref{st:results}, two types of distributions are presented: (1) instantaneous forces (per unit length), obtained by the integration of the surface stress maps along the dorso-ventral axis, at 5 selected time points covering half a period of oscillation, and (2) period-averaged stress maps, where the normal stress is averaged either in absolute value or with its sign. In each experiment, the flow field is collected over three flapping periods and 30 pairs of velocity fields (10/period) are selected for the evaluation of hydrodynamic forces. Instantaneous distributions are averaged over 6 equivalent time frames (considering the symmetry between the left and right strokes). The period averaged distributions are based on the 30 time frames. The instantaneous force (per unit length) is decomposed into the $x$ and $y$ directions, yielding the thrust $F_x$ and lateral force $F_y$ acting on the fin. The thrust force corresponds to the useful power spent by the fin (when it points in the negative $x$ direction, propelling the system upstream). The lateral forces correspond to the wasted power. Hence, we define the efficiency ratio as the power invested by the fin into useful thrust divided by the total rate of work done on the fluid:

\begin{equation}
\eta = \frac{\abs{\langle F_x \rangle} \cdot \langle \overline{u}_{fin} \rangle}{\abs{\langle F_x \rangle} \cdot \langle \overline{u}_{fin} \rangle + \langle \abs{F_y} \rangle \cdot 2Af}
\label{eq:efficiency}
\end{equation}

This definition is equivalent to the Froude efficiency: it measures the ability of the fin to convert work into upstream propulsion \citep{Eloy2013, Quinn2015}. The over lines indicate a spatial average and the brackets indicate a time average over a period. Note that $F_x$ is averaged with its sign; the absolute value is applied after averaging, whereas $F_y$ is averaged in absolute value directly because the fin is wasting energy due to lateral forces from both sides. The fluid velocity is averaged over the foil virtual boundary rather than over the whole volume, in order to better capture the specific ability of each hydrofoil to generate downstream fluid motion (see table \ref{t:foil_hydrodynamics}). This classification based on $\eta$ is compared to the usual efficiency classification based on the Strouhal number, where the window of efficient propulsion is considered to be [0.2,0.4] \citep{Triantafyllou2000, Taylor2003}.

Many authors have calculated the denominator of equation \ref{eq:efficiency} for pitching fins (input power) as the period-averaged product of torque and angular velocity at the base (Dewey 2013, Quinn 2015, Lucas 2015, Egan 2016, David 2017, Floryan 2017, Lucas 2017, Rosic 2017, Zhu 2017). This efficiency metric is appropriate for hydrodynamic experiments where forces and torque sensors are placed at the attachment rod of the flapping propulsor. In the present study, no load cells are involved and all information about the forces imparted by the foil on the fluid are extracted solely from the flow velocity field. In this context, a definition of $\eta$ based on local force components integrated on the fin surfaces is more suitable. A similar calculation using surface integration of hydrodynamic stresses was used in \citet{Liu2017}.

The uncertainties on the pressure and force distributions can be obtained using error propagation arguments and an analysis of the inaccuracy on the Lagrangian path reconstruction (used in the material acceleration evaluation). The uncertainties on the particles positions in our PTV experimental set up are $\sigma_x=\sigma_y \simeq 3.6 \mu \textrm{m}$ and $\sigma_z \simeq$ 32 $\mu$m, which imply velocity uncertainties of $\sigma_{u_x}=\sigma_{u_y} \simeq$ 0.002 m/s and $\sigma_{u_z} \simeq$ 0.018 m/s. Noise propagation from the velocity field to the material acceleration and to the integrated pressure field has been the object of many studies \citep{Liu2006, Violato2011, deKat2013, vanOudheusden2013, Wang2017}. We can derive an expression for the pressure uncertainty, relevant for the present calculation scheme. As a first step, we express the material acceleration uncertainty (with $a_i=Du_i/Dt$) as a sum of the error propagated from the velocity field (first term under the square root) and the uncertainty from the Lagrangian path line reconstruction (second term):

\begin{equation}
\sigma_{a_i} = \sqrt{\frac{2\sigma_{u_i}^2}{\Delta t^2} + \big( (\vec{\sigma}_u \cdot \vec{\nabla})u_i \big)^2}
\label{eq:error_material_acceleration}
\end{equation}

The pressure field is the result of a spatial integration, its uncertainty therefore depends on the spatial resolution $\Delta x_i$ and on the number of nodes $n$ crossed along the integration path. Moreover, the pressure integration algorithm involves a median polling among a collection of $N$=8 paths, which reduces the uncertainty further by a factor of $\sqrt{\frac{\pi}{2(N-1)}}$ \citep{Kenney1962}. Because any of the $x$, $y$ or $z$ directions can be followed by the integration paths, the estimated pressure uncertainty is based on the average of errors in all three directions \citep{Dagenais2019}:

\begin{equation}
\sigma_p = \frac{\rho}{3} \sqrt{\frac{\pi}{14}}\sum_{i=1}^{3}n \Delta x_i \sqrt{\frac{2\sigma_{u_i}^2}{\Delta t ^2}
+\big( (\vec{\sigma}_u \cdot \vec{\nabla})u_i \big)^2}
\label{eq:error_pressure}
\end{equation}

The local pressure uncertainties are propagated to the force uncertainties using classic error propagation through an integration step, and the resulting values are shown with error bars in section \ref{st:results}. Due to temporal and spatial averaging, these uncertainties are reduced by an additional factor of $\sqrt{N_{time} \times N_{spatial}}$. $N_{time}$ is the number of time frames employed in the period average (6 or 30 for the instantaneous and period averaged distributions, respectively). $N_{spatial}$ is 1 for the instantaneous distributions and 20 in the case of the period averaged curves of figure \ref{f:ave_normal_force}, which are spatially averaged over the left/right and dorsal/ventral symmetric sides of the fin, as well as over 5 dorso-ventral rows for each curve.

\section{Results}
\label{st:results}

\subsection{Hydrofoils kinematics}
\label{sst:results_kinematics}

\begin{figure}
\centering
\includegraphics[width=13cm, trim={2cm 0 0 0}]{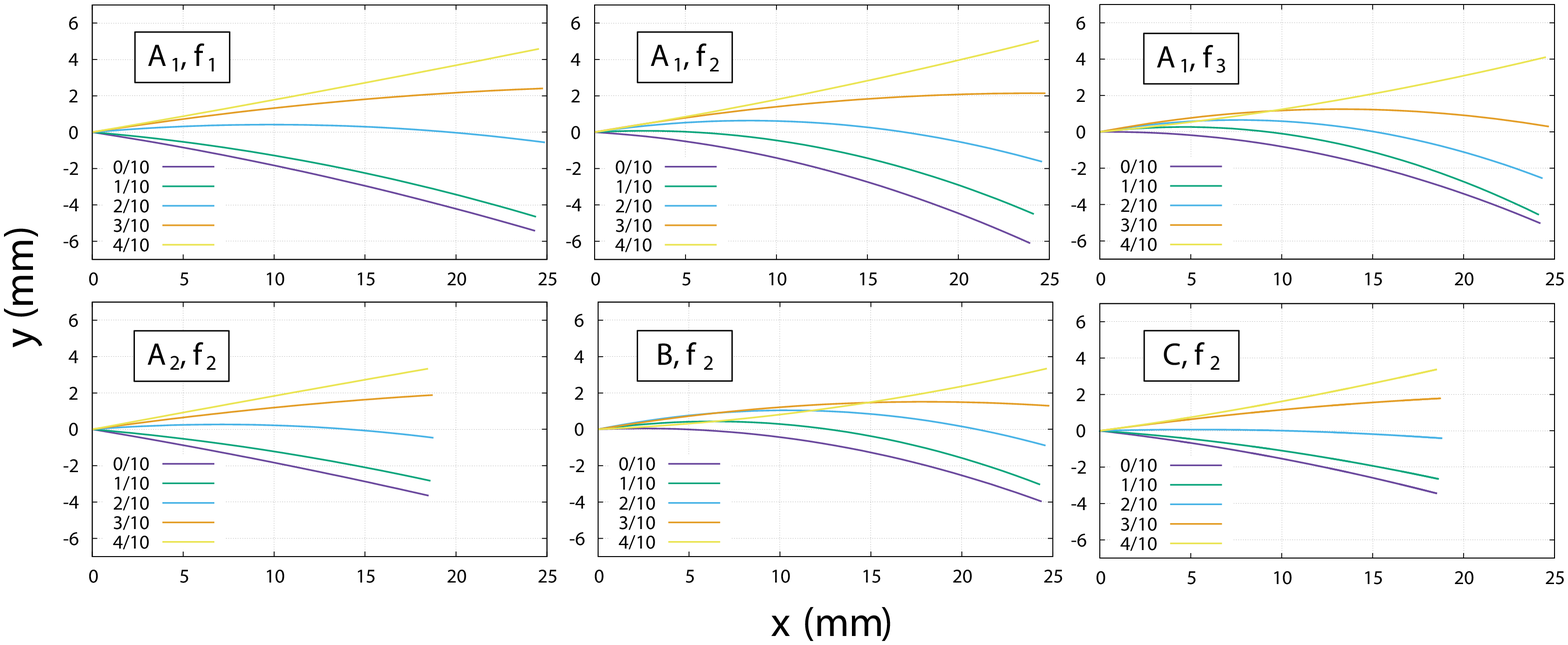}
\caption{Hydrofoils midlines at 5 time points equidistant over half a period.}
\label{f:midlines}
\end{figure}

The geometry and flapping frequency directly affect the fluid-structure interactions of the hydrofoils. The deflection profiles of the midlines along the course of a flapping cycle are shown in figure \ref{f:midlines}. The midlines excursions are used to evaluate $A$, the tip amplitude, which is needed to calculate $St$. The phase lag between the peduncle and the tip ($\Delta \phi$) is determined by measuring the angle between the fin midline and the horizontal plane (at the peduncle and at the tip) at regular time points during the motion cycle. The results are listed in table \ref{t:foil_hydrodynamics}. The largest phase lag is found for the wide fin geometry (shape B), with $\Delta \phi$ twice as large as for shape A\textsubscript{1}. This tip recoil is associated to a smaller excursion amplitude at the tip ($A$). Increasing the flapping frequency of shape A\textsubscript{1} induces a larger phase lag, but in this case, the diminution of the tip amplitude is less important. Almost no phase lag is found for the shorter fins, independently of the trailing edge geometry (shapes A\textsubscript{2} and C). The shorter geometries thus behave more like rigid fins. The capacity of each hydrofoil to generate streamwise fluid velocity can be analyzed by averaging the value of $u_x$ over multiple periods and the whole volume ($\overline{u}_{vol}$) or over the fins virtual boundary ($\overline{u}_{fin}$), the latter option displaying more obvious differences (see table \ref{t:foil_hydrodynamics}). The highest streamwise velocity is produced by the bilobed short fin (shape C), followed by the long narrow fin flapping at maximal rate (shape A\textsubscript{1} at $f\textsubscript{3}$). The slowest fluid velocity is observed in the case of the wider fin (shape B) followed by the long narrow fin flapping at minimal rate (shape A\textsubscript{1} at $f\textsubscript{1}$). Based on the Strouhal numbers alone, we would anticipate that the foil shape A\textsubscript{1} with flapping frequencies $f\textsubscript{2}$ and $f\textsubscript{3}$ lie outside of the propulsive efficiency range [0.2,0.4]. Nevertheless, the efficiency ratio $\eta$ offers a different perspective on that classification, as will be shown in section \ref{sst:results_force_dist}.

\begin{table}
\begin{center}
\begin{tabular}{c c c c c c c c c c}
shape                   & frequency                &  $\Delta \phi$ &  $A$ & $\overline{u}_{vol}$  & $\overline{u}_{fin}$  & $St$  & $\eta$ \\
                             &                                 &      ($T$)        & (mm) &  (mm/s)                    & (mm/s)                     &          &            \\
\hline
A\textsubscript{1} & $f\textsubscript{1}$ & 0.11              &  11.0  &   52.6                         &     48.3                     & 0.37   & 0.14   \\
A\textsubscript{1} & $f\textsubscript{2}$ & 0.18              &  12.0  &   54.5                        &      54.8                     & 0.61   & 0.17   \\
A\textsubscript{1} & $f\textsubscript{3}$ &  0.32              &  10.3  &   56.5                        &     58.4                     & 0.69   & 0.17   \\
A\textsubscript{2} & $f\textsubscript{2}$ &  0.08              &   8.0   &   55.3                        &      53.8                    & 0.40   & 0.13   \\
B                           & $f\textsubscript{2}$ &  0.36              &  7.7   &   46.4                         &      47.4                    & 0.39   & 0.19   \\
C                          & $f\textsubscript{2}$ &   0.07             &  7.5    &   56.7                         &      61.4                    & 0.38   & 0.10
\end{tabular}
\caption{Kinematics and flow regime of the hydrofoils: phase lag between peduncle and tip angles ($\Delta \phi$, in fraction of a period), total amplitude covered by the fin tip over a period ($A$), streamwise velocity component averaged over three periods ($\overline{u}_{vol}$, averaged over the whole volume, and $\overline{u}_{fin}$, averaged over the virtual surface enclosing the fin), Strouhal number ($St$) and efficiency ratio ($\eta$).}
\label{t:foil_hydrodynamics}
\end{center}
\end{table}

\subsection{Spatio-temporal distributions of fluid forces}
\label{sst:results_force_dist}

\begin{figure}
\centering
\includegraphics[width=13cm, trim={2cm 0 0 0}]{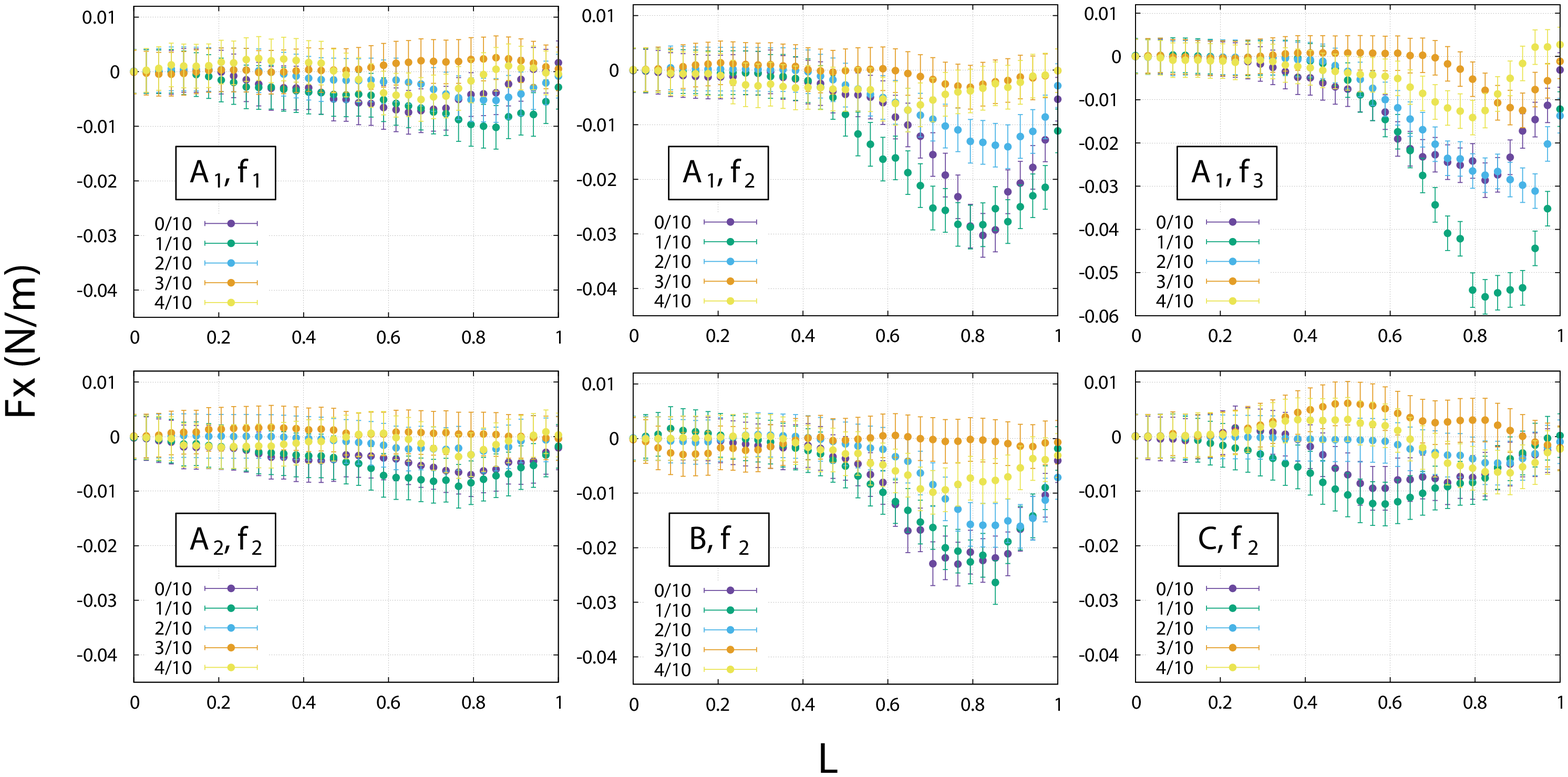}
\caption{Thrust ($x$-component) force per unit length acting on the fin (in N/m), at 5 time points equidistant over half a period.}
\label{f:thrust}
\end{figure}

\begin{figure}
\centering
\includegraphics[width=13cm, trim={2cm 0 0 0}]{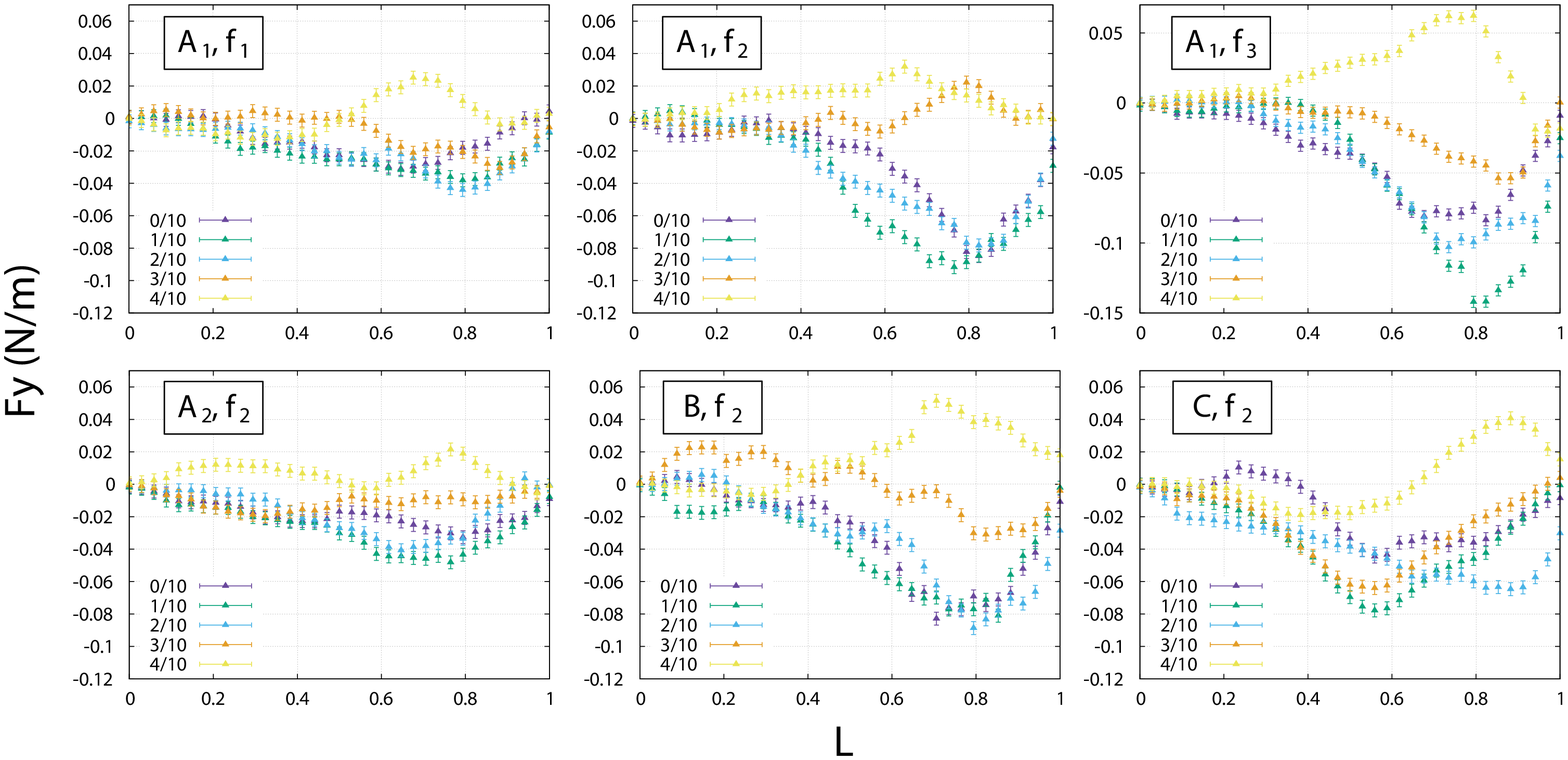}
\caption{Lateral ($y$-component) force per unit length acting on the fin (in N/m), at 5 time points equidistant over half a period.}
\label{f:lateral_force}
\end{figure}

\begin{figure}
\centering
\includegraphics[width=13cm, trim={2cm 0 0 0}]{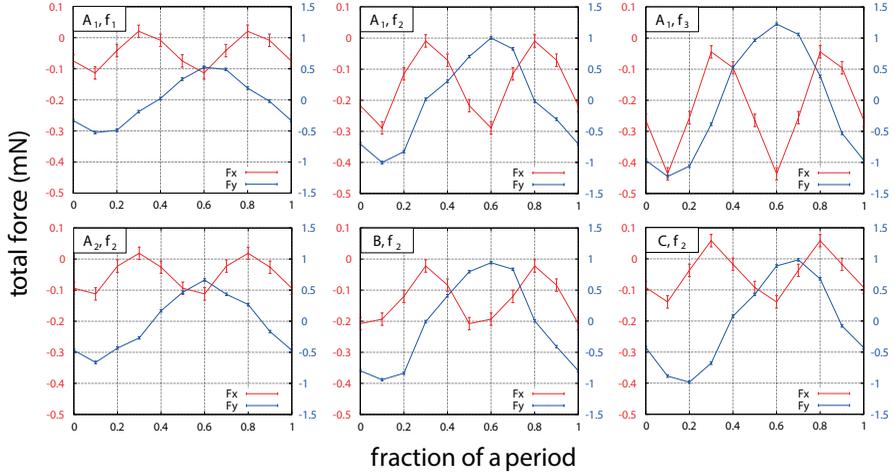}
\caption{Total forces (in mN) acting on the fin, at 5 time points equidistant over half a period (thrust and lateral forces, in the $x$ and $y$ directions respectively).}
\label{f:total_forces}
\end{figure}

\begin{figure}
\centering
\includegraphics[width=13cm, trim={2cm 0 0 0}]{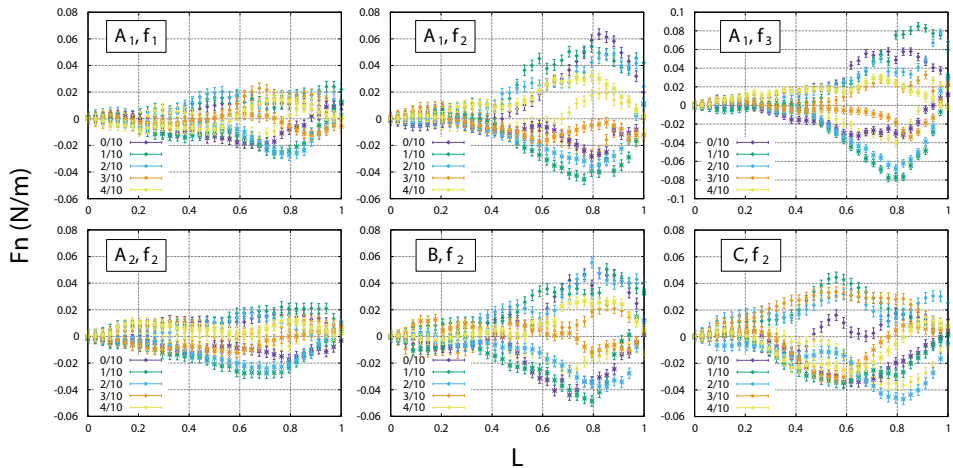}
\caption{Normal force per unit length (in N/m), acting on both sides of the fin (left side: diamonds, right side: stars) at 5 time points equidistant over half a period.}
\label{f:normal_force}
\end{figure}

The color code of figures \ref{f:thrust}, \ref{f:lateral_force} and \ref{f:normal_force} follows that of figure \ref{f:midlines} in terms of time partition. Figure \ref{f:thrust} illustrates the distributions of fluid forces (per unit length) in the $x$ direction, acting on the fins, along their proximo-distal axis. Half a period is represented using five equidistant time points (naturally, the opposite half-period would produce symmetrical distributions). Negative values correspond to useful propulsive thrust as the system is pushed upstream. In all cases, the maximal amount of thrust is produced when the tip is leaving its maximal point and initiating its motion back towards the center (time point 1/10 in green). The thrust then reduces gradually until time point 3/10 (orange), after which it increases again as the fin approaches the opposite extremity of its excursion (time point 4/10 in yellow). The maximal amount of thrust is generated by the long fin at maximal flapping rate (shape A\textsubscript{1} at $f\textsubscript{3}$), followed by that same geometry at frequency $f\textsubscript{2}$. The wider fin (shape B) also produces significant thrust although slightly lower than the narrow geometry. For these three cases, the maximal amount of force is generated by the most distal portion of the fin, peaking close to 80\% of the total length. The thrust is drastically reduced in the case of the shorter fin (shape A\textsubscript{2}). In comparison, the short fin with bilobed edge (shape C) produces slightly larger amount of thrust at time points 0/10 and 1/10, but this is compensated by a positive distribution of $F_x$ at time 3/10, detrimental to propulsion. In that specific case, the maximum amount of force is generated closer to the center of the foil. The time evolution of the total force in the $x$ direction (integrated over the whole fin surface) is shown in red in figure \ref{f:total_forces}. Because of the symmetry between the left and right strokes of the fin, maximum thrust is generated twice per cycle.

Figure \ref{f:lateral_force} presents the $y$ component of the hydrodynamic force (per unit length) acting on the fin. This lateral force corresponds to wasted energy (not useful to propulsion). The most thrust-producing foils (shapes A\textsubscript{1} at $f\textsubscript{3}$, A\textsubscript{1} at $f\textsubscript{2}$ and shape B) are also the ones where the most energy in lost in accelerating fluid in the lateral direction. Moreover, the distribution on $F_y$ is correlated to the distribution of $F_x$ both temporally and spatially, with a maximum lateral force produced at time points 0/10, 1/10 and 2/10, at a position close to 80\% of the fin length. The time points which are typically not associated to thrust production (3/10 and 4/10) correspond to lower lateral forces. The bilobed geometry (shape C) presents a more complex spatio-temporal distribution of $F_y$, where the force peak alternates between two locations during the half period: at 50\% of the fin length for time points 1/10 and 3/10, and 80\% for time points 2/10 and 4/10. The lowest levels of wasted energy due to lateral forces are found for the long fin at low frequency (shape A\textsubscript{1} at $f\textsubscript{1}$), followed by the shorter fin (shape A\textsubscript{2}). The time evolution of the total lateral force is shown in blue in figure \ref{f:total_forces}. The left and right strokes produce mirroring force profiles due to the motion symmetry.

\begin{figure}
\centering
\includegraphics[width=13cm, trim={2cm 0 0 0}]{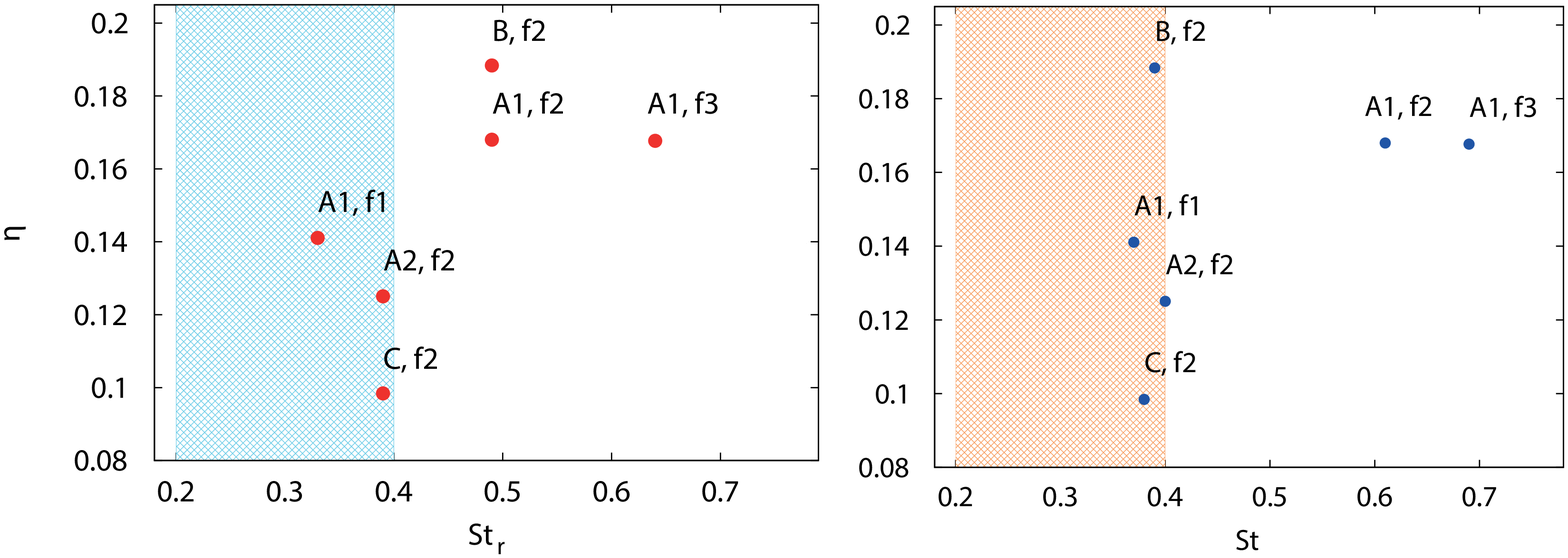}
\caption{Efficiency ratio (as defined by equation \ref{eq:efficiency}), as a function of the \textit{rigid} Strouhal number $St_r$, calculated based on the rigid projection of the peduncle (left) and the \textit{flexible} Strouhal number $St$, calculated based on the tip excursion amplitude of the flexible foils (right).}
\label{f:efficiency}
\end{figure}

As a measure of propulsive energy, we use $\eta$, the ratio between the energy employed to produce thrust and the sum of that energy with the energy employed to produce lateral forces (equation \ref{eq:efficiency}). Figure \ref{f:efficiency} allows to compare the efficiency thus defined for all foils with their respective flapping rates, as a function of $St_r$ and $St$. \citet{David2017} have raised the question of whether or not the Strouhal number calculated from the tip excursion amplitude is a good metric for the width of a vortex wake. They showed that for highly flexible foils, the width of the wake is overestimated by the large tip excursion, and that a rigid projection of the pitching peduncle approximates better the vortices spacing. Interestingly, the fact that flexible fins have larger tip excursion amplitudes than their rigid counterparts is not verified in all our experiments. It is true for geometries A\textsubscript{1} and A\textsubscript{2}, but not for shapes B and C, where the amplitude is smaller than the rigid projection. Therefore, in some cases, the flexibility of a fin tends to decrease its excursion amplitude and its Strouhal number.

The most efficient shape is the wider foil (B), followed by the long narrow shape (A\textsubscript{1}), which is equally efficient at frequencies $f\textsubscript{2}$ and $f\textsubscript{3}$ but drops in efficiency at lower frequency ($f\textsubscript{1}$). The shorter fins are the less efficient ones, with the lowest value of $\eta$ found for the short fin with a bilobed edge (C). This is consistent with the fact that flexible fins are typically more efficient than their rigid counterparts \citep{Dewey2013}. Indeed, shapes A\textsubscript{2} and C, because of their specific geometries, behave almost like rigid panels, as can be inferred from the close to zero phase lag between the peduncle and the tip angles (table \ref{t:foil_hydrodynamics}), and from the visualization of their midlines (figure \ref{f:midlines}).

In figure \ref{f:efficiency}, the dashed regions denote the range of Strouhal numbers typically associated to efficient thrust production. Interestingly, the classification based on the definition of $\eta$ shows that shape A\textsubscript{1} at frequencies $f\textsubscript{2}$ and $f\textsubscript{3}$ is more efficient than shapes A\textsubscript{2} and C, even though their Strouhal numbers lie outside of the usual window of efficiency [0.2,0.4]. In all cases, the efficiency ratio $\eta$ remains below 0.2, indicating that less that 20\% of the input power is used for propelling the system in the upstream direction, which is not particularly efficient.

\begin{figure}
\centering
\includegraphics[width=13cm, trim={2cm 0 0 0}]{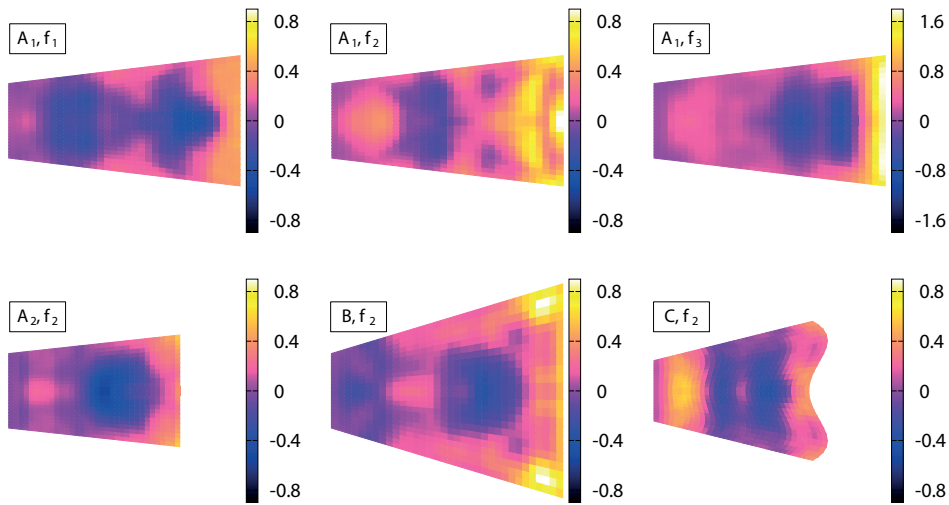}
\caption{Period averaged signed value of the normal stress acting on the sides of the fin (in Pa).}
\label{f:tension}
\end{figure}

\begin{figure}
\centering
\includegraphics[width=13cm, trim={2cm 0 0 0}]{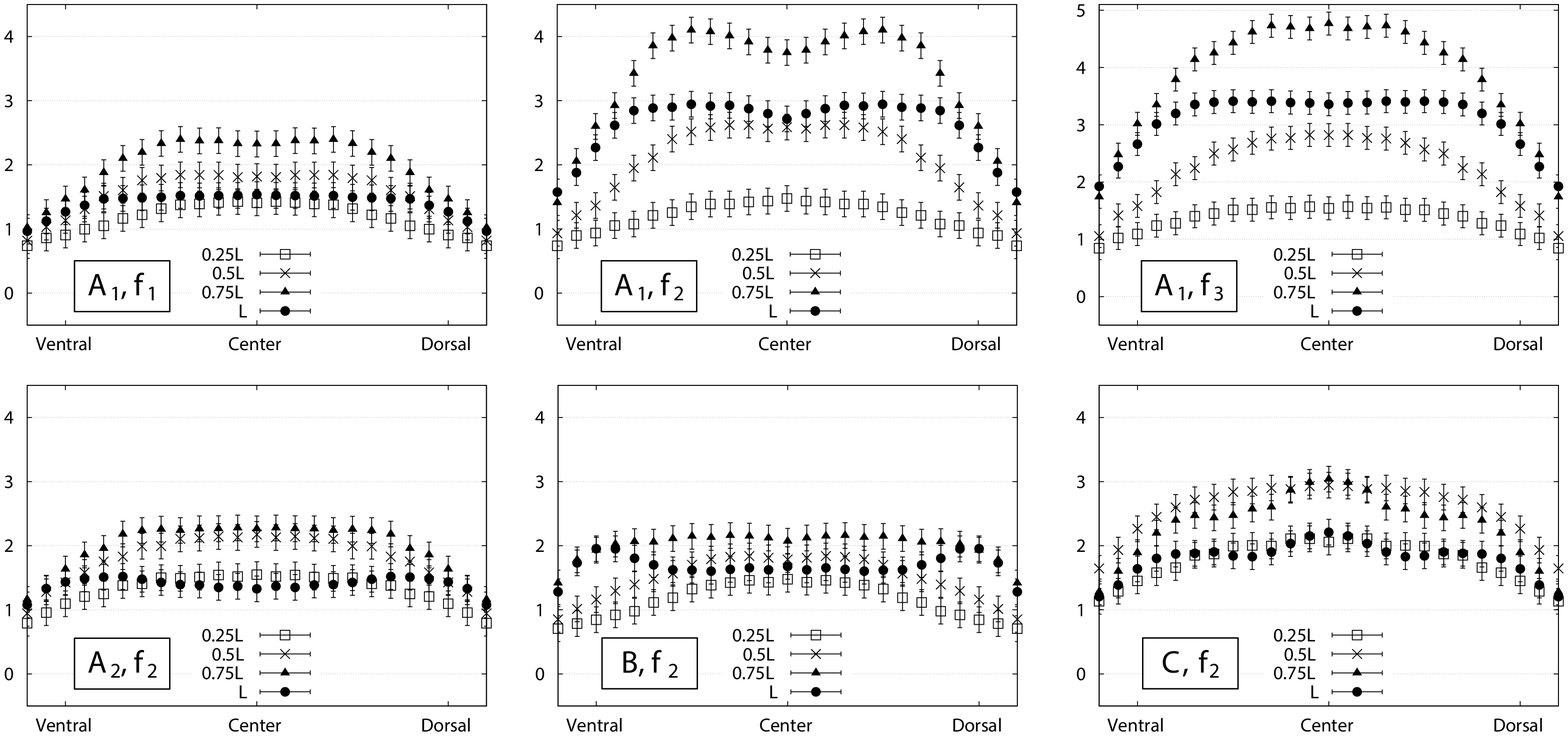}
\caption{Period averaged absolute value of the normal stress acting on the sides of the fin (in Pa, $\sigma_p$=0.2 Pa).}
\label{f:ave_normal_force}
\end{figure}

The normal force distributions (per unit length) on both sides the hydrofoils are illustrated in figure \ref{f:normal_force}. Low flapping rate (shape A\textsubscript{1} at $f\textsubscript{1}$) or shorter fin length (shape A\textsubscript{2}) result in lower amount of normal forces across both surfaces, peaking close to 70\% of the total fin length. At intermediate or higher flapping rate (shape A\textsubscript{1} at $f\textsubscript{2}$ or $f\textsubscript{3}$), when the fin is wider (shape B), or when the edge is bilobed (shape C), the normal forces increase and the maximum magnitude is shifted close to 80\% of the fin. Moreover, the asymmetry between the left and right sides of the hydrofoils becomes more important. The magnitudes are asymmetric, and in the most extreme cases, the normal force has the same sign on both fin sides (see for example time points 3/10 and 4/10 for shape A\textsubscript{1} at $f\textsubscript{2}$). These asymmetries translate into larger internal tension within the material. If the sum of the forces on both sides is positive, the material inside the fin is subjected to an outward stretching as it experiences a net force in the direction of the outward normal. Contrarily, if the sum on both sides is negative, the resultant force points towards the inside of the fin, which is then subjected to a net compression. A perfectly symmetric force distribution on both sides (sum equal to zero) would indicate that one side is pulled while the other is pushed with the exact same magnitude, therefore, the hydrodynamic stress would result in the acceleration of the foil, with no internal stress caused by the net external contact forces. To evaluate the spatial distribution of internal tension across the hydrofoils, the normal stress is averaged over a full cycle of flapping, with its sign. The results are presented as color maps in figure \ref{f:tension}. Negative values denote an excess of compression over the oscillation cycle, positive values denote stretching. All membranes experience an overall stretching over the tip region. The most extreme occurrence of that effect is found at the tip of shape A\textsubscript{1} at $f\textsubscript{3}$. A high concentration of stretching is also seen for shape A\textsubscript{1} at $f\textsubscript{2}$, localized in the very center of the tip, and for shape B, where the concentration zone is shifted to the corners. Additionally, there is a compression zone located distally to the center of the foils, between approximately 50\% and 75\% of the total length. This is seen in all cases except for shape A\textsubscript{1} at $f\textsubscript{2}$, where the internal compression is relaxed in the center of the fin. Furthermore, the normal stress distributions are averaged over a period, in absolute values, revealing which portions of the fin are most prominently involved in the production of hydrodynamic forces. These spatial distributions are presented in figure \ref{f:ave_normal_force}, with stress curves covering the dorso-ventral axes of the fin and taken at different locations across its length (25\%, 50\%, 75\% and 100\%). The maximum values are always found at $\sim$75\% of the foil length. The variations across the surface are marked mostly in the case of the long and narrow geometry (shape A\textsubscript{1}), where the curves at 0.25$L$, 0.5$L$, 0.75$L$ and $L$ become more and more separated as the flapping frequency increases. It is noteworthy that a special stress pattern emerges for shape A\textsubscript{1} at $f\textsubscript{2}$, where the curve at 0.75$L$ adopts a bilobed signature, not observed for $f\textsubscript{1}$ nor $f\textsubscript{3}$. The excess of normal stress on the tip corners of shape B is also visible in the corresponding graph (stress curve at $L$). Finally, all fin geometries and flapping rates result in lower levels of hydrodynamic forces generated at the lateral edges (dorsal and ventral) of the hydrofoils.

\subsection{Vorticity fields and wake structures}
\label{sst:results_vorticity}

\begin{figure}
\centering
\includegraphics[width=13cm, trim={2cm 0 0 0}]{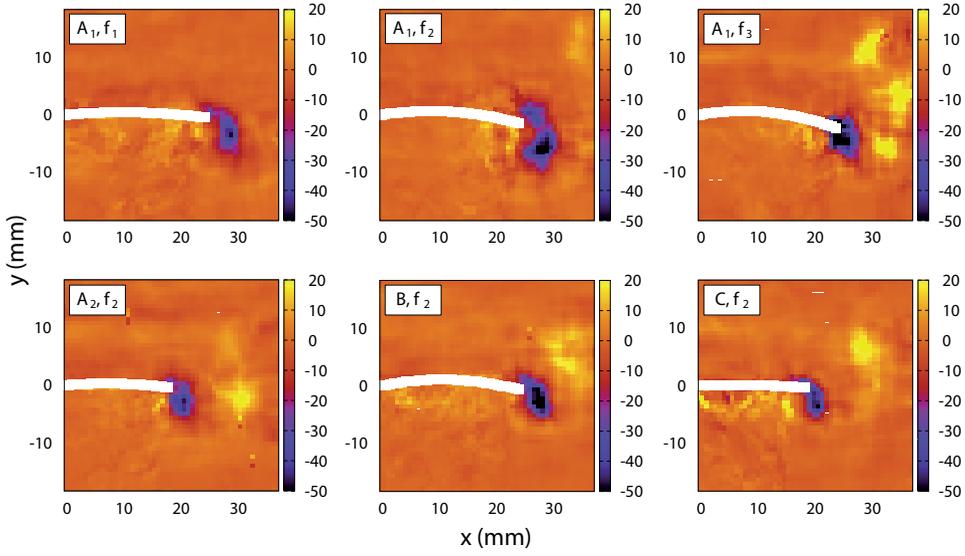}
\caption{Snapshots of $\omega_z$, the $z$-component of the vorticity field (in s\textsuperscript{-1}) in the plane intersecting the fins half-way through their span, as they are crossing the $y=0$ axis, travelling towards positive $y$.}
\label{f:vorticityZ_angle0}
\end{figure}

\begin{figure}
\centering
\includegraphics[width=13cm, trim={2cm 0 0 0}]{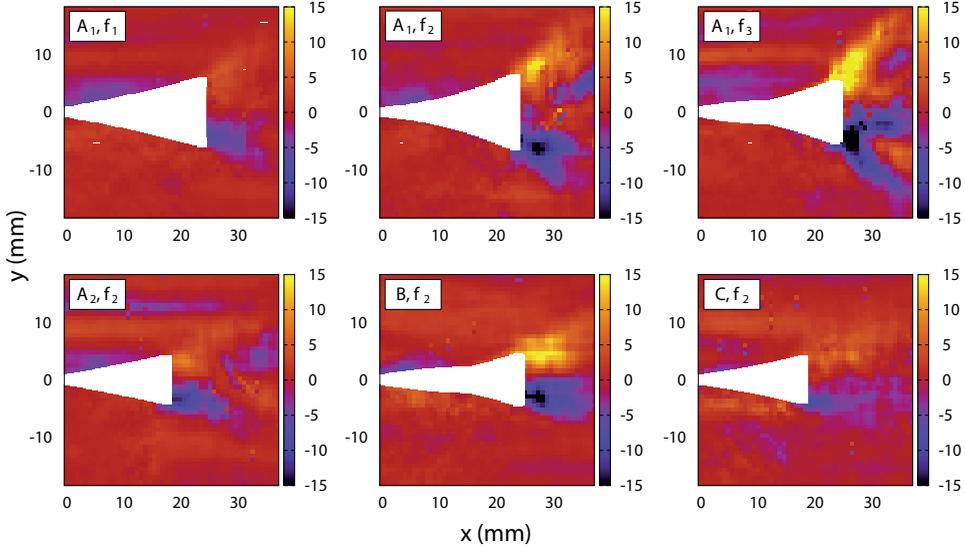}
\caption{Period-average of $\omega_z$, the $z$-component of the vorticity field (in s\textsuperscript{-1}) in the plane intersecting the fins half-way through their span, with the cycle of motion envelopes indicated by white areas.}
\label{f:vorticityZ_average}
\end{figure}

The vorticity field ($\vec{\omega}=\vec{\nabla} \times \vec{u}$) is very informative about characteristic flow structures, revealing relations between wake topology, surface distributions of fluid forces and propulsive efficiency. To allow comparison with previous work, we present the distributions of $\omega_z$, the $z$ component of the vorticity, in 2D planes intersecting the fins at mid-span. Instantaneous $\omega_z$ maps are shown in figure \ref{f:vorticityZ_angle0}, as the foils are crossing the $y=0$ axis (corresponding to 2/10 of a period, color-coded in pale blue in figures \ref{f:midlines}, \ref{f:thrust}, \ref{f:lateral_force} and \ref{f:normal_force}). Period-averaged $\omega_z$ maps are shown in figure \ref{f:vorticityZ_average}.

During the flapping cycle, the instant when the fins cross the $y=0$ axis occurs between the points of maximal and minimal thrust production (see figure \ref{f:total_forces}). Figure \ref{f:vorticityZ_angle0} illustrates that a vortex with a core of negative $\omega_z$ (rotating clockwise) forms at the tip of each fin. For shape A\textsubscript{1} at frequency $f\textsubscript{1}$, the vortex is almost detached at this point, whereas at $f\textsubscript{2}$, it is fully detached, with a second smaller vortex forming at the tip. The size and strength of these clockwise vortices correlate with propulsive efficiency: the largest vortices, with maximal $\abs{\omega_z}$ of the order of 50 s\textsuperscript{-1}, are found for shape A\textsubscript{1} at $f\textsubscript{2}$, shape A\textsubscript{1} at $f\textsubscript{3}$ and shape B at $f\textsubscript{2}$, the three cases with maximal $\eta$ coefficients. Counterclockwise vortices are also present in the snapshot vorticity fields for shapes A\textsubscript{1} (at $f\textsubscript{3}$), A\textsubscript{2}, B and C. They are located more downstream as they have been shed during the precedent half-cycle of motion. In the lower frequency cases for shape A\textsubscript{1} ($f\textsubscript{1}$ and $f\textsubscript{2}$), they are not discernable in the field of view, most likely because they have been shed downstream past $x=35$ mm by that time. The persistence in the close downstream wake of these positive $\omega_z$ vortices do not appear to be correlated with propulsive efficiency, nor their magnitude or orientation. Rather, it is the flow structures directly adjacent to the tip which embody the capacity of a fin to generate efficient thrust. It is interesting to observe that the curvature of the trailing edge has an impact on the position at which the vortices are being shed. Indeed, counter-rotating vortices are added to the wake along the $y=0$ axis for a straight edge geometry (shape A\textsubscript{2}), whereas they are detaching at higher $y$ positions for a bilobed trailing edge (shape C).

The period-averaged $z$-vorticity also presents a correlation with propulsive efficiency: the three hydrofoils with highest $\eta$ coefficients (shape A\textsubscript{1} at $f\textsubscript{2}$ and $f\textsubscript{3}$ and shape B at $f\textsubscript{2}$) display the highest means, around 15 s\textsuperscript{-1} (figure \ref{f:vorticityZ_average}). For all shapes and frequencies, the period-averaged $\omega_z$ indicates that a surplus of clockwise vorticity is induced at the tip when the fin travels in the $y<0$ region, and vice-versa. In other words, the fluid tends to rotate more towards the exterior of the motion envelope close to the tip. In the wake surrounding more proximal portions of the foils, the opposite effect is observed: the period-averaged $z$-vorticity is slightly positive in the $y<0$ region, and vice-versa. This phenomenon is linked to the period-averaged surface maps of normal stress (figure \ref{f:tension}), denoting an excess of stretching close to the tip, and a surplus of compression towards the center and more proximal surface regions. Moreover, stronger stretching force at the tip appears to be related to higher mean vorticity close to the trailing edge, as seen for shape A\textsubscript{1} at $f\textsubscript{2}$ and $f\textsubscript{3}$ and shape B at $f\textsubscript{2}$. From figure \ref{f:vorticityZ_average}, an estimation of the vortex shedding angle can be drawn, corresponding to the opening angle of the maximal $\abs{\omega_z}$ zone. This angle is wider in the cases of shape A\textsubscript{1} at $f\textsubscript{2}$ and $f\textsubscript{3}$, roughly following the tip curvature of the motion envelopes. The most efficient fin (shape B) presents a less pronounced shedding angle, which remains aligned with the motion envelope outline. The shedding angle seems particularly constrained for the two shorter fins behaving more like rigid panels (shapes A\textsubscript{2} and C), which are also the least thrust-efficient hydrofoils. In those cases, the high vorticity areas are almost aligned with the $y=0$ axis. When comparing the period-averaged wake signatures of shape A\textsubscript{1} at $f\textsubscript{1}$, $f\textsubscript{2}$ and $f\textsubscript{3}$, we observe that increasing the flapping frequency results in the appearance of a reversed circulation area in the downstream wake, close to $x=33$ mm, where $\omega_z$ changes its sign. This mean vorticity reversal does not seem related to propulsive efficiency, since it is also slightly visible in the case of shape A\textsubscript{2} at $f\textsubscript{2}$, one of the least efficient foils in terms of thrust production.

\section{Discussion and outlook}
\label{st:discussion}

We have established a systematic experimental approach to compare the hydrodynamic stress maps on different fin shapes with varying flapping frequency, allowing us to identify the effects of each parameter on the spatio-temporal distributions of fluid forces, and link them to the direct wake topology. In our analysis, we payed special attention to the ability of each hydrofoil to generate thrust (force propelling the system upstream), lateral forces (not contributing to propulsion) and the resulting propulsion efficiency. We also investigated the spatial distributions of external contact forces normal to the fins surfaces and the corresponding internal tension accumulating within the material. We have found that the long narrow geometry generates the most thrust (at equal frequency), and that increasing the flapping frequency increases thrust. Between both short geometries, the bilobed trailing edge allows to produce slightly more thrust than the straight edge, but only temporarily during the period. At a certain point during the flapping cycle, it is also pushed downstream by the fluid (the opposite of propulsive thrust). It is typically the distal region of the fins which participates more in thrust generation, except for the short bilobed geometry whose central portion is the one most involved. As for the lateral forces, detrimental to propulsive efficiency, the short straight geometry (at equal frequency) reaches the lowest levels. For the long geometry, reducing the flapping frequency also reduces the lateral force magnitude throughout the cycle.

The examination of instantaneous and period-averaged vorticity fields yielded the conclusion that the formation of larger and stronger vortices directly at the tip as well as the accumulation of excess vorticity close to the trailing edge over a full period are related to better propulsive efficiency. Among all cases tested, the most efficient geometry was found to be the long wide fin, followed by the long narrow one, whose efficiency remained similar at higher frequency but dropped significantly at the lowest one. The least efficient fin morphology is the short fin with bilobed trailing edge. This type of tip curvature is ubiquitous in nature. It is interesting, both for the study of biological swimmers and for the design of artificial fins, to discover that the naturally widespread bilobed shape does not improve efficiency nor does it grant more thrust producing capabilities to a simple fin model at Reynolds numbers which are on the low side of the inertial regime. It is more likely that an authentic fish fin with a bilobed trailing edge is coupled with specific elasticity profiles and complex controlled motion programs, in order to achieve its functional role. It could also be that the advantages of this tip curvature appear only for a specific combination of parameters such as length and flapping frequency. Besides, we have observed that the bilobed geometry results in a dynamical shift of fluid force peaks along the fin length during the stroke cycle. This more complex dynamics of force distributions could be used actively by an appendage to achieve fine maneuvers. The external contact forces from the fluid acting normally on both foils surfaces result in an accumulation of internal tension throughout the flapping cycle. The fin tip is subjected to an excess of tissue stretching, an effect which becomes more important as the foil becomes longer, or as it flaps faster. The stretching tension concentrates towards the tip corners for a wider geometry. Moreover, the foil membranes experience an overall compression zone at about two thirds of their total length. Interestingly, in the case of the long narrow geometry, flapping at frequency $f\textsubscript{2}$, this compression region partly disappears. In that specific case, a particular stress pattern emerges from 75\% of the total length up to the tip, where the normal stress (averaged in absolute value over a period) presents a double peak signature. For all fin geometries, the period averaged absolute value of the normal stress reaches its maximum at about 75\% of the total length. These stress maps are very indicative of how the fin architecture must be conceived in order to withstand specific concentrations of mechanical loads and internal tension throughout its oscillation cycle.

The flexible tails of rayed-fins fishes display an impressively wide range of morphological traits and provide a great model system to study the relation between fin geometry and propulsive efficiency \citep{Affleck1950, McNeill1974, Lauder1989, Weihs1989, Lauder2000, Lauder2007a, Blake2009, Flammang2011b}. Complex combinations of environmental constraints participate in the selective evolution of fish fins, and the resulting elastic properties, geometry and kinematics of each specie constitute an elegant compromise between competing survival demands such as the necessity to achieve fine control of the fin surface for delicate maneuvering, to generate quick acceleration bursts or to maintain long distance efficiency \citep{Webb1975, Lauder1983, Lauder1989, Lauder2006, Lauder2015}. It is tempting to assume that evolution has optimized the geometric and kinematic parameters of fins to meet with the functional demands of distinct habitats. However, drawing a parallel between form and function requires taking into account the evolutionary history of fins morphologies, and considering not only the external shape, but also internal structures, in order to avoid mistaking correlations between morphology, habitat and function for causal relationships \citep{Lauder1983, Lauder1989}. Previous investigations on the interplay between shape and stiffness of fin models and their capacity to generate thrust efficiently have already demonstrated that efficiency varies in a complex manner as a function of the different fin attributes, and that it is not a simple matter to identify an optimal shape \citep{Feilich2015}. Rather, efficiency can be maximized by several combinations of parameters, depending on the flow conditions, and it remains a great challenge to draw general conclusions about the functional advantages of geometrical and kinematic aspects. The present work illustrates how the investigation of hydrodynamic stresses based on three-dimensional PTV experiments can provide precious information about the mechanical interplay between a fin-like structure and the surrounding fluid. It paves the way for future experimental studies using direct evaluation of fluid forces on the surface of submerged flexible appendages, allowing to tackle directly the fundamental problem of form and function in the context of propulsion through a fluid medium. To conclude, the possibility to resolve the 3D instantaneous hydrodynamic stress fields on small-scaled models of deformable structures opens the door to many applications in the engineering design of aeronautic and aquatic vehicles or the field of biophysics concerned with animal locomotion.

\section{Acknowledgements}
This work was funded by the Swiss National Science Foundation (SNF) via a Sinergia research grant as well as a UZH Forschungskredit Candoc grant. We are grateful for interdisciplinary discussions with Tinri Aegerter, Anna Ja\'zwi\'nska, Ivica Kicic, Sahil Puri and Siddhartha Verma. We are very thankful to J.O. Dabiri \textit{et al.} for making the \textit{queen2} pressure algorithm available \cite{Dabiri2014}. Declaration of Interests: The authors report no conflict of interest.

\bibliographystyle{jfm}
\bibliography{stress_fin_physics_bib}

\end{document}